\documentclass[a4paper,11pt]{article}
\usepackage[utf8]{inputenc}
\usepackage{jinstpub} 
\usepackage{float}
\usepackage{threeparttable}
\usepackage{booktabs}
\usepackage{multirow}
\usepackage{tabularx}
\usepackage{graphicx}     % for \includegraphics
\usepackage{subcaption}
\usepackage{soul}

\usepackage{lineno}
\usepackage{xcolor}
\title{\boldmath A compact, low-power epithermal neutron counter for lunar water detection
}

\author[1, 7]{Julian Cuevas-Zepeda}
\author[2]{Phoenix Alpine}
\author[3]{Brenda A. Cervantes-Vergara}
\author[3]{Claudio Chavez}
\author[1,3,4,7]{Juan Estrada}
\author[5]{Erez Etzion}
\author[1,3]{Guillermo Fernandez-Moroni}
\author[3, 7]{Nathan Saffold}
\author[6]{Miguel Sofo-Haro}
\author[3]{Javier Tiffenberg}

\affiliation[1]{Department of Astronomy and Astrophysics, University of Chicago, Chicago, IL 60637, USA}
\affiliation[2]{Department of Aerospace Engineering, University of Illinois Urbana–Champaign, Urbana, IL 61801, USA}
\affiliation[3]{Fermi National Accelerator Laboratory, Batavia, IL 60510, USA}
\affiliation[4]{Instrumentation Division, Brookhaven National Laboratory, Upton, NY 11973, USA}
\affiliation[5]{School of Physics and Astronomy, Tel Aviv University, Tel Aviv 69978, Israel}
\affiliation[6]{Comisión Nacional de Energía Atómica (CNEA) y Consejo Nacional de Investigaciones Científicas y Técnicas (CONICET), Universidad Nacional de Córdoba, Córdoba 5000, Argentina}
\affiliation[7]{Kavli Institute for Cosmological Physics, University of Chicago, Chicago, IL 60637, USA}

\emailAdd{juliancz@uchicago.edu}

\abstract{

The detection and characterization of lunar water are critical for enabling sustainable human and robotic exploration of the Moon. Orbital neutron spectrometers, such as instruments on Lunar Prospector and the Lunar Reconnaissance Orbiter, have revealed hydrogen-rich regions near the poles but are limited by coarse spatial resolution and low counting efficiency. 
We present a compact, lightweight, and low-power epithermal neutron detector based on boron-coated silicon imagers, designed to probe subsurface hydrogen at decimeter scales from mobile platforms such as lunar rovers. This instrument leverages the high neutron capture cross-section of $^{10}$B to convert epithermal neutrons into detectable $\alpha$ and $^{7}$Li ions in a fully-depleted silicon imager, providing a unique event topology to identify neutrons while suppressing backgrounds. Monte Carlo simulations demonstrate that a 3\,$\mu$m boron layer achieves optimal neutron detection efficiency, further enhanced with polyethylene moderation to improve sensitivity to the 0.4\,eV--500\,keV epithermal energy range. For a 10\,cm$^2$ active area, the detector achieves sensitivity to H$_2$O weight fractions as low as 0.01\,wt\% in a 15\,minute measurement. This scalable, portable, low-mass design is well-suited for integration into upcoming Artemis and commercial lunar rovers, providing a transformative capability for in-situ resource prospecting and ground-truth validation of orbital measurements.}

\begin{document}

\maketitle
\flushbottom

\section{Introduction}

Lunar resource exploration is a key objective in modern lunar science and will help plan sustainable human and robotic missions on the Moon. Water is a critical resource that could be used by future missions for life-support systems, rocket propellant, radiation shielding, and irrigation, among other potential applications. The presence of water and other volatile compounds on the Moon has been speculated upon since the 1950s~\cite{Urey_1952,Watson_1961}, with the development of cosmochemical models to predict lunar elemental abundances validated by spectroscopic observations of moonlight. The Apollo missions provided lunar soil samples that enabled the first direct measurements of the Moon's water content.
While initial spectrometric analyses of the Apollo lunar soil samples found that the Moon was anhydrous, modern re-analyses with more advanced technology have detected water at the tens of ppm level~\cite{saal_volatile_2008}, which motivated a flurry of follow-up measurements.

Currently, it is widely recognized that water is present on the Moon, but the precise abundance, distribution, and accessibility of this water remain only partially understood.
Previous lunar missions, including Clementine~\cite{Clementine_1996}, Lunar Prospector~\cite{Feldman_1998}, Chandrayaan-1~\cite{Chandrayaan_2009}, and the Lunar Reconnaissance Orbiter (LRO)~\cite{LRO_2010}, have revealed compelling evidence of hydrogen-rich regions near the lunar poles, indicating potential reservoirs of water ice within permanently shadowed regions (PSRs)~\cite{LCROSS_2010}. However, these regions have only been identified on $\mathcal{O}(\mathrm{10\,km})$ length scales, necessitating the development of more sensitive, high-resolution detection techniques to create a detailed map. We note that, in the context of lunar resources, hydrogen and water are often conflated because they are both volatile and readily escape the thin lunar atmosphere. In contrast, oxygen is abundant in the silicate-rich lunar soil. Therefore, the presence of hydrogen provides a strong indication of water, and furthermore, water could be produced on the Moon by harvesting hydrogen and hydroxyl-rich regions.~\cite{PROSPECT_2024}.

Current water mapping strategies rely primarily on three types of instruments: neutron spectrometers, which measure hydrogen-induced suppression of epithermal and fast neutrons; infrared spectrometers, which identify hydroxyl and water absorption features on the lunar surface; and radar instruments, which probe subsurface ice through its dielectric properties.
Among these techniques, neutron spectroscopy has proven to be uniquely capable of inferring bulk hydrogen content several tens of centimeters beneath the regolith. Instruments such as the Lunar Exploration Neutron Detector (LEND) on LRO and the Lunar Prospector Neutron Spectrometer (LP-NS) have provided global maps of the Moon's hydrogen distribution, but they are limited by coarse spatial resolution and counting statistics constrained by detector efficiency and mass. These limitations motivate the development of compact, high-efficiency neutron detection systems that are suitable for future lunar exploration missions.

In this work, we present a novel approach to detect lunar water using boron-coated silicon sensors. This technology leverages the high thermal neutron capture cross-section of $^{10}$B, which produces charged reaction
products ($\alpha$ particles and $^7$Li) that can be directly detected with low-noise silicon sensors. This design integrates a boron conversion layer with a scientific-grade silicon imaging sensor, providing a lightweight, compact, and low-power neutron detector with sensitivity extending to epithermal neutron energies.
Unlike conventional $^3$He or scintillator-based detectors, this boron-coated silicon sensor offers high-resolution imaging capabilities with background discrimination in a scalable package, enabling deployment on CubeSats, lunar landers, and lunar rovers.

This paper discusses the motivation for high-resolution water mapping of the Moon, reviews the status of neutron detection instruments, and presents the boron-coated silicon imager as a competitive solution for next-generation lunar resource exploration. 
Sections~\ref{sec:DetConcept} and~\ref{sec:MoonWaterNeutrons} introduce the concept of detecting epithermal neutrons with boron-coated silicon imagers and epithermal neutrons as probes of lunar water. Sections~\ref{sec:sims} and~\ref{sec:rover} present detailed simulations of sensor performance, estimate instrument sensitivity, and outline a benchmark deployment concept on a lunar rover. Section~\ref{sec:lowpower} presents a conceptual design, including expected power, mass, and data budget. Finally, Section~\ref{sec:conclusion} presents the conclusions and outlook.

\section{Boron-coated silicon imagers for neutron detection}
\label{sec:DetConcept}

\begin{figure}[t]
    \centering
    \includegraphics[width=1\linewidth,angle=0]{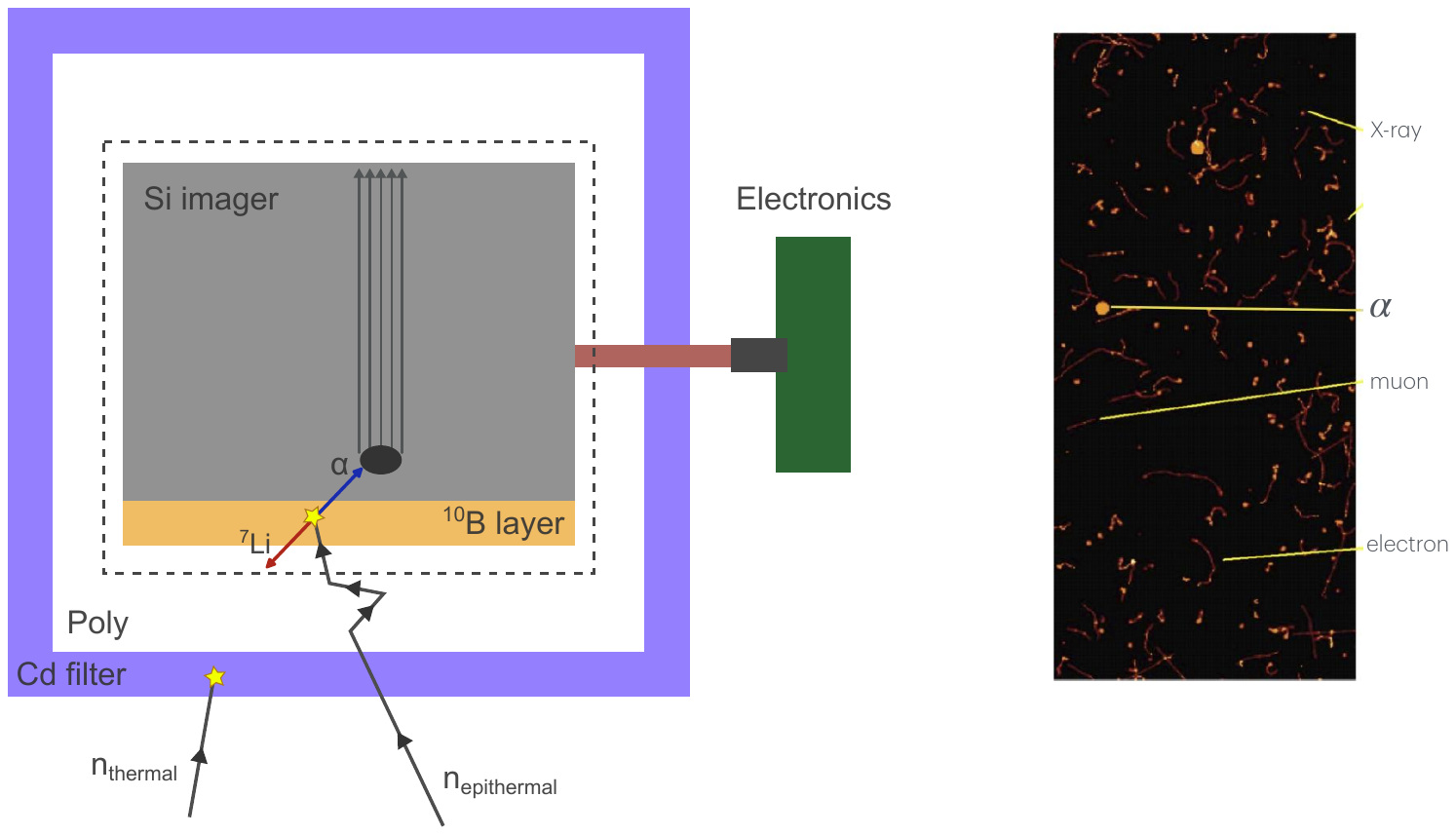}
    \caption{Left: Schematic of epithermal neutron detection with boron-coated silicon imaging sensor. An incident neutron is captured in a $^{10}$B layer producing an $\alpha$ particle and a $^7$Li ion. When these charged particles reach the silicon, they produce a very dense ionization cloud. The resulting charge is collected by the silicon sensor electrodes. Right: Example image showing charge clusters from various particle interactions and their unique event topologies. The straight tracks result from highly energetic cosmic rays (ie. muons, protons), the serpentine tracks correspond to particles with multiple small-angle deflections (ie. betas), the point-like hits can be attributed to photon interactions (ie. X-rays), and the large circular hits correspond to alpha particles that deposit all of their energy in the sensor and produce the plasma effect.}
    \label{fig:neutroncapture}
\end{figure}

The boron-coated silicon detector technology, originally developed for thermal and ultracold neutron measurements~\cite{Blostein2014, Kuk_2021}, can be directly applied to the detection and imaging of epithermal neutrons (0.4\,eV-500\,keV). The detection mechanism relies on the neutron capture reaction in $^{10}$B: $^{10}\mathrm{B}(n,\alpha_0)^{7}\mathrm{Li}\quad (6\%) \quad 
\text{and} \quad ^{10}\mathrm{B}(n,\alpha_1\gamma)^{7}\mathrm{Li}\quad (94\%),$ where an incident neutron is absorbed in the $^{10}$B layer, producing an $\alpha$ particle and a $^7$Li ion emitted back-to-back.
The  $\alpha$ particle and Li ion are produced with $\sim$1\,MeV kinetic energies, and therefore are highly ionizing with ranges of only a few micrometers in solids. When the boron layer is deposited directly onto the surface of a fully depleted silicon detector, a fraction of the $\alpha$ particles and Li ions escape the boron and enter the silicon bulk, where they produce 
a dense column of electron-hole pairs. This column is considered a plasma if the Debye length is small compared to its dimensions \cite{Estrada_Plasma_2011}. In this scenario, the charge distribution produced in the silicon detector is clearly distinguishable from other types of radiation, providing a powerful tool for background suppression.

The ionization mechanism, illustrated in Figure~\ref{fig:neutroncapture}, is described in three steps:

\begin{itemize}
\item{Neutron Capture in Boron:} An incident epithermal neutron is absorbed in the thin $^{10}$B layer. The $^{10}$B(n,$\alpha$)$^7$Li reaction converts the neutral particle into two charged particles.
\item{Emission and Escape of Reaction Products:} The $\alpha$ and $^7$Li ions are emitted in opposite directions. If emitted toward the silicon detector, and if their residual energy after passing through the boron layer exceeds the energy loss in any silicon dead layer, they penetrate the active silicon volume. For this application, 
we will use backside processing to reduce the silicon dead layer to $<$100\,nm, as is standard for UV sensors.
\item{Charge Collection in Silicon:}
As the ions traverse the fully depleted silicon, they generate electron–hole pairs along a short track. The high local ionization density produces the “plasma effect”~\cite{Estrada_Plasma_2011}, in which the charge cloud expands laterally before collection, creating a characteristic circular distribution of charge. In typical silicon imagers such as scientific CCDs and CMOS sensors, this charge is collected in neighboring pixels of the pixel array. The resulting contiguous pixels form a charge cluster in the imager with topology that is unique to the incident particle. The total collected charge is proportional to the particle energy, enabling particle identification and discrimination against gamma or cosmic-ray events.
\end{itemize}

The probability of detecting a neutron event depends on two competing effects: the likelihood of neutron capture in the $^{10}$B layer (enhanced by thicker coatings) and the escape probability of the $\alpha$ and $^7$Li ions (reduced by thicker coatings). Optimal thicknesses for thermal neutrons are typically around 1-2\,$\mu$m \cite{Blostein2014}, whereas for epithermal neutrons, thicker layers or multiple boron coatings can increase absorption probability, taking advantage of the 1/$v$ behavior of the $^{10}$B cross section at low energies (where $v$ is the velocity of the neutron).

A silicon imager coated with boron provides three key features for detecting epithermal neutrons. First, it enables efficient conversion of neutrons to ions in the \(^{10}\text{B}\) layer. Second, the detector allows for the direct detection of charged reaction products in fully-depleted silicon, providing sub-pixel ($<$15\,$\mu$m) position resolution as discussed in Ref.~\cite{Kuk_2021}. Lastly, it offers a unique event topology and energy discrimination, which help to reject background events.

Imaging with thermal neutrons captured on a borated CCD was demonstrated in Ref.~\cite{Blostein2014}, and a similar CCD-based approach was used for imaging with ultracold neutrons at Los Alamos~\cite{Kuk_2021}. Neutron detection with commercial CMOS detectors was successfully demonstrated in Ref.~\cite{Perez_2018,2021_cmos_nano}. In this context, we propose to leverage these developments to devise a solid-state epithermal neutron counter compatible with deployment on mobile platforms in the space environment. This detector concept will allow the design of instrumentation for planetary neutron spectroscopy, particularly relevant for identifying hydrogen-rich regions that indicate the presence of H$_2$O.

\section{Epithermal neutrons as probes for water on the Lunar surface} 
\label{sec:MoonWaterNeutrons}
Neutron spectroscopy is a proven technique for measuring planetary hydrogen abundances~\cite{LRO_2010}. High-energy neutrons are produced via spallation when Galactic cosmic rays strike the planetary surface. These neutrons gradually lose energy through scattering in the regolith, and as they approach thermal energies their capture probability increases, reducing the emergent surface flux.
Since hydrogen nuclei are similar in mass to neutrons, hydrogen is very efficient at moderating neutrons compared to heavier nuclei. The presence of hydrogen in the lunar soil suppresses the emergent flux of epithermal neutrons, and neutrons with energies between $\sim$0.4\,eV up to $\sim$500\,keV are most sensitive to hydrogen and its abundance variations. Therefore, measurements of a suppressed epithermal neutron flux from the lunar surface are indicative of imprints of hydrogen and/or deposits of water ice. Previous experiments mapping hydrogen on the Moon with neutron spectroscopy have reported a noticeable suppression of epithermal neutrons at high latitudes around the lunar poles~\cite{Feldman_1998,LRO_2010}, predominantly within permanently shadowed regions (PSRs).

\begin{figure}[t]
    \centering
    \includegraphics[width=0.49\linewidth]{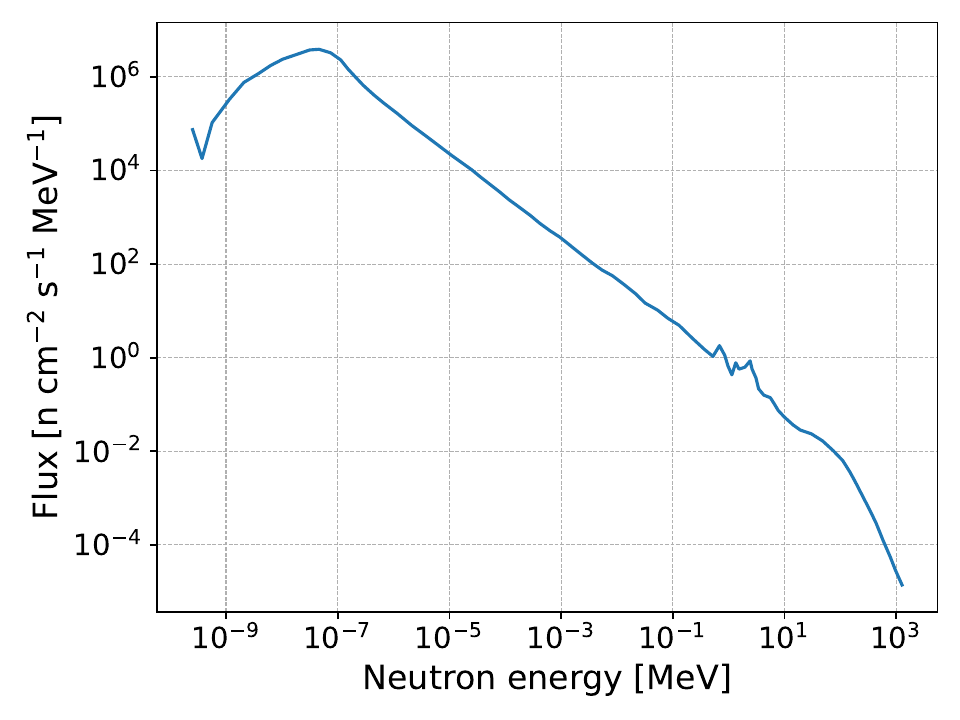}
    \caption{Neutron flux energy spectra for a soil composition representative of that expected at the lunar poles, digitized from Ref.~\cite{Lawrence:2006}.} \label{fig:neutronfluxmoon}
\end{figure}

\begin{figure}[ht!]
    \centering
    \includegraphics[width=0.49\linewidth]{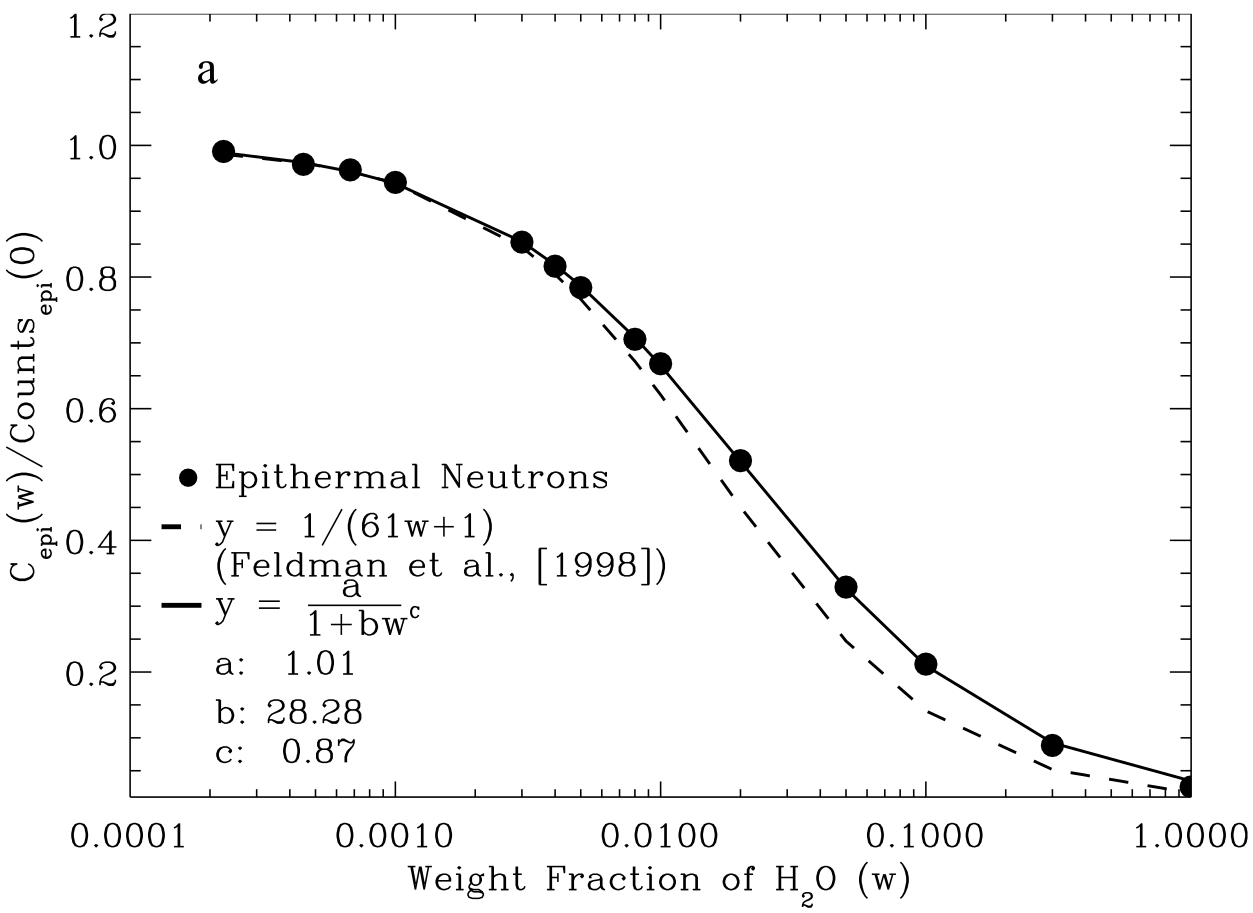}
    \includegraphics[width=.49\linewidth]{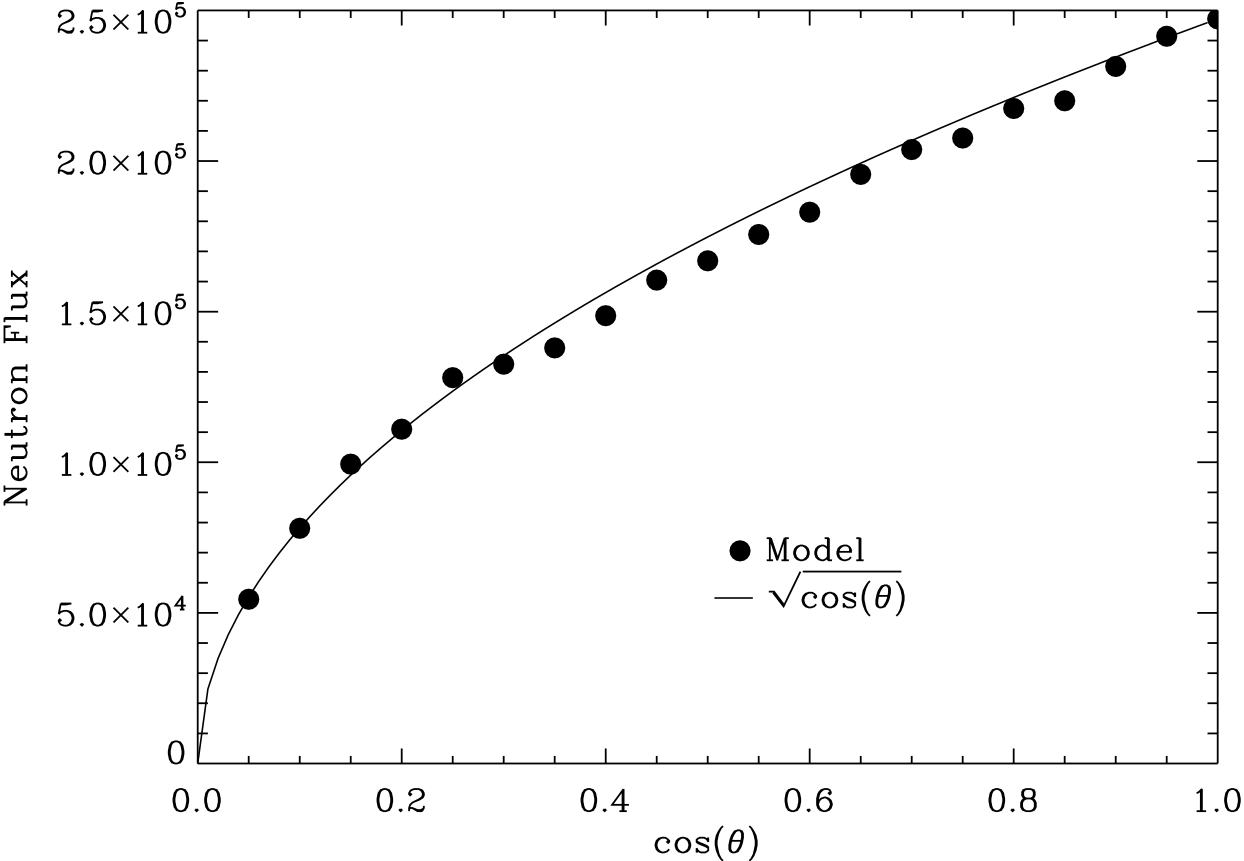}
    \caption{Figures from Ref.~\cite{Lawrence:2006}. Left: Ratio of wet to dry epithermal neutron counts as a function of H$_2$O weight fraction $w$, assuming a soil composition representative of that expected at the lunar poles, and its fit with Eq.~\eqref{eq:nfluxwwater}. Right: Angular distribution of the modeled neutron flux for the same soil type. The solid line is $\sqrt{\cos{\theta}}$, normalized to the $\cos{\theta}=1$ point.} \label{fig:n_water_angles}
\end{figure}

To quantify abundances via neutron spectroscopy, experimental data are compared with models of the expected neutron rates. For our initial study, we use the epithermal neutron flux at the lunar surface reported by LP-NS~\cite{Lawrence:2006} and shown in Figure~\ref{fig:neutronfluxmoon}, assuming a soil composition representative of that expected at the lunar poles. Ref.~\cite{Lawrence:2006} also provides an empirical relation for the epithermal neutron counts, $C_{\rm epi}$, as a function of the H$_2$O weight fraction, $w$, for a model with a single-layer regolith:
\begin{equation} \label{eq:nfluxwwater}
    C_{\rm epi}(w)=C_{\rm epi}(w=0)\frac{a}{1+bw^c}\,,
\end{equation}
where the fit parameters are $a=1.01$, $b=28.28$, and $c=0.87$. Figure~\ref{fig:n_water_angles} (left) shows the ratio of wet to dry epithermal neutron counts as a function of $w$, taken from Ref.~\cite{Lawrence:2006}. Finally, to estimate our detection efficiency, we use the neutron angular distribution from the same work, shown in Figure~\ref{fig:n_water_angles} (right).

Epithermal neutron flux is relatively stable under varying environmental conditions~\cite{Lawrence:2006}. In particular, temperature-dependent variations affect only the parameter $a$ in Eq.~\eqref{eq:nfluxwwater}, with changes up to 3\% between 100\,K and 400\,K. Additionally, in stratified models where hydrogen-enriched soil is buried beneath a dry layer, the epithermal neutron flux remains relatively constant for dry layer thicknesses up to $\sim$50\,g\,cm$^{-2}$. In contrast, the thermal neutron flux shows a much stronger dependence on dry soil thickness in these layered models, making thermal neutrons a potentially useful probe of subsurface structure. However, thermal neutrons are more sensitive to temperature variations than the epithermal component. Therefore, the epithermal component of the neutron flux provides a more reliable measure of the lunar soil's water content.
\begin{figure}[ht!]
    \centering
    \includegraphics[width=.49\linewidth]{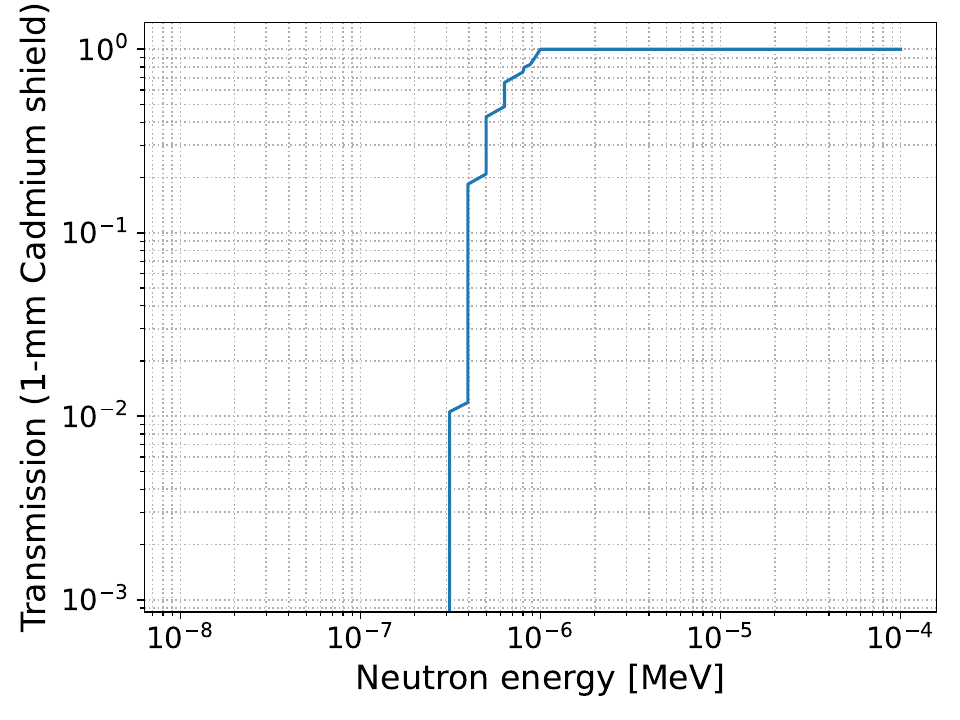}
    \caption{\label{fig:filter} Neutron transmission through a 1\,mm layer of cadmium, computed from data in Ref.~\cite{DMELLOW2007690}. The Cd filter strongly suppresses thermal neutrons, providing more than three orders of magnitude attenuation below 0.2\,eV, while maintaining nearly 100\% transmission for neutrons with energies above 1\,eV.} 
\end{figure}

To enhance sensitivity to the epithermal neutron flux, we suppress the background from thermal neutrons with a high-pass neutron filter in the detector assembly. For instance, a 1\,mm cadmium (Cd) filter effectively suppresses neutrons with energies below 1\,eV, exhibiting a sharp turn-on curve as shown in Figure~\ref{fig:filter}. By introducing a Cd filter, the acquired data is most sensitive to the epithermal component of the neutron flux, which is more robust against temperature fluctuations and provides the best indication of the water.

\section{Modeling the Performance of Boron-Coated Silicon Imagers on the Lunar Surface}
\label{sec:sims}

\subsection{Silicon as a neutron detector}

In general, silicon sensors are poor neutron detectors. The usual detection mechanism in silicon imagers involves ionization produced from charged particles or photon absorption/scattering processes that directly transfer most of the deposited energy into electron-hole pair generation. As electrically neutral particles, neutrons only produce ionization in silicon through direct scattering with silicon nuclei~\cite{Lindhard1963, Chavarria2016}. In nuclear scattering, energy is dominantly dissipated via vibrations in the crystalline lattice rather than ionization, making neutron scatters in silicon hard to identify with ionization-based detectors. 
Therefore, given the low energy response of neutrons in silicon coupled with the low interaction cross-section \cite{PhysRev.126.1105} between neutron and silicon, silicon detector systems have gone vastly ignored as effective neutron detectors. However, based on the results in Refs.~\cite{Blostein2014, Kuk_2021} discussed in Section~\ref{sec:DetConcept}, we can enhance the sensitivity of silicon imagers to epithermal neutrons through the addition of a thin layer of $^{10}$B, a neutron moderator, and a high-pass neutron filter (see Figure~\ref{fig:filter}).

\begin{figure} [ht!]
    \centering
    \includegraphics[width=0.7\linewidth]{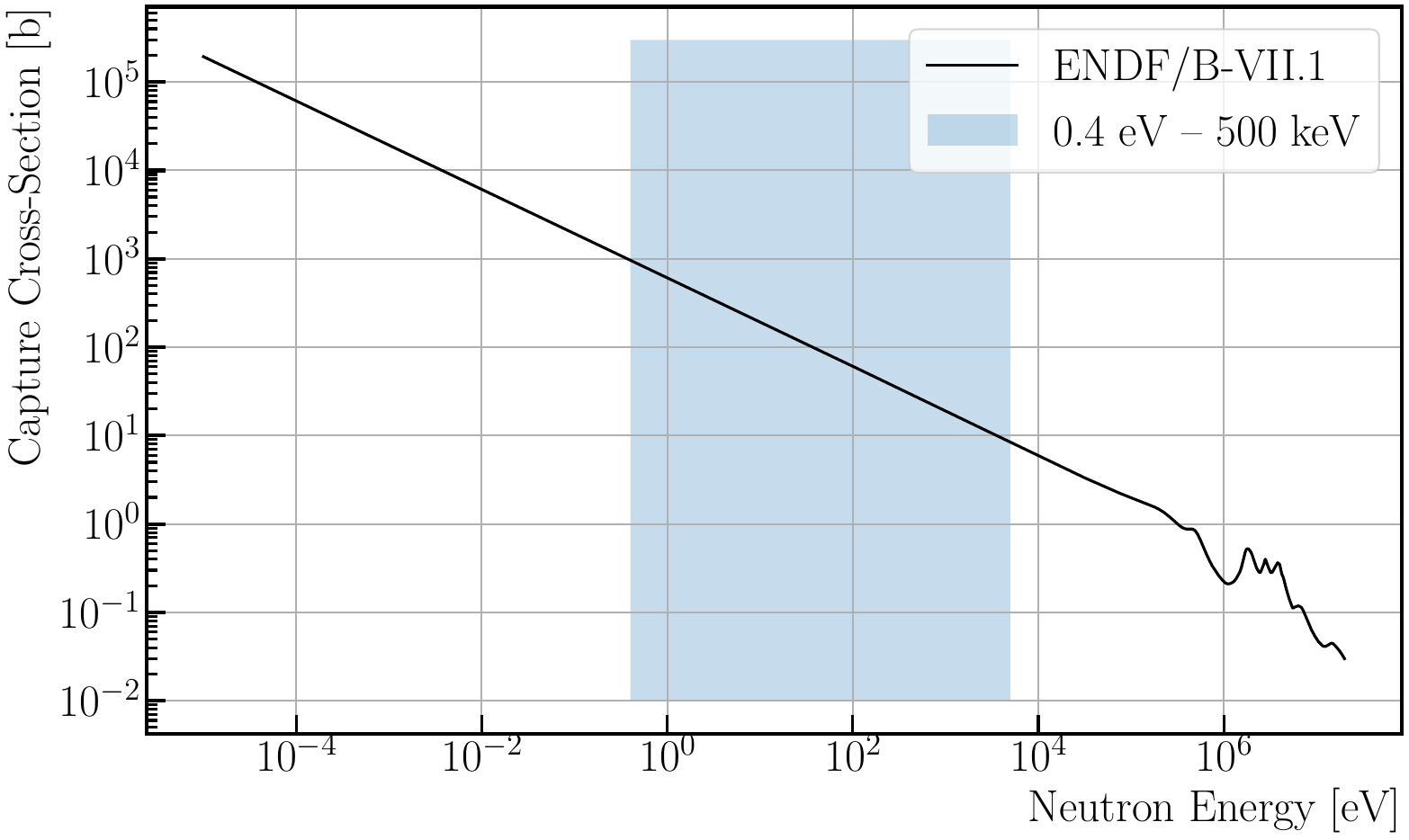}
    \caption{Evaluated neutron capture cross-section for the $^{10}$B(n, $\alpha_i$)$^7$Li reaction. The shaded region represents the range of epithermal neutrons of interest for our study.}
    \label{fig:cap-cs}
\end{figure}

To assess the response of the boron-coated silicon sensor to epithermal neutrons, a series of simulations were launched with MCNPX \cite{pelowitz2011mcnpx}. We model a 750\,$\mu$m silicon sensor coated with a highly-enriched boron film ($>$96\% purity $^{10}$B) under various conditions. The $^{10}$B(n, $\alpha_i$)$^7$Li capture reaction is modeled via the ENDF/B-VII.1 \cite{chadwick2011endfb7, hale2006b10endf} neutron cross-section libraries and shown in Figure~\ref{fig:cap-cs}. When inspecting the energy dependence of the neutron capture cross-section, we note that for increasing neutron energies there is a decreasing probability of capture. Our initial study aims to 1) determine the optimal boron-film thickness for maximum ion detection in our silicon sensor, 2) determine the optimal moderator thickness to enhance epithermal neutron capture in $^{10}$B, and 3) estimate the signal response of a silicon sensor to neutrons from the lunar surface.

\subsection{Neutron-Induced Ion Detection Efficiency in Silicon}

\subsubsection{Film-Thickness Selection}
\label{sec:filmthickness}

\begin{figure}[htb!]
  \centering
  % First subfigure
  \begin{subfigure}[b]{0.49\linewidth}
    \centering
    \includegraphics[width=0.9\linewidth]{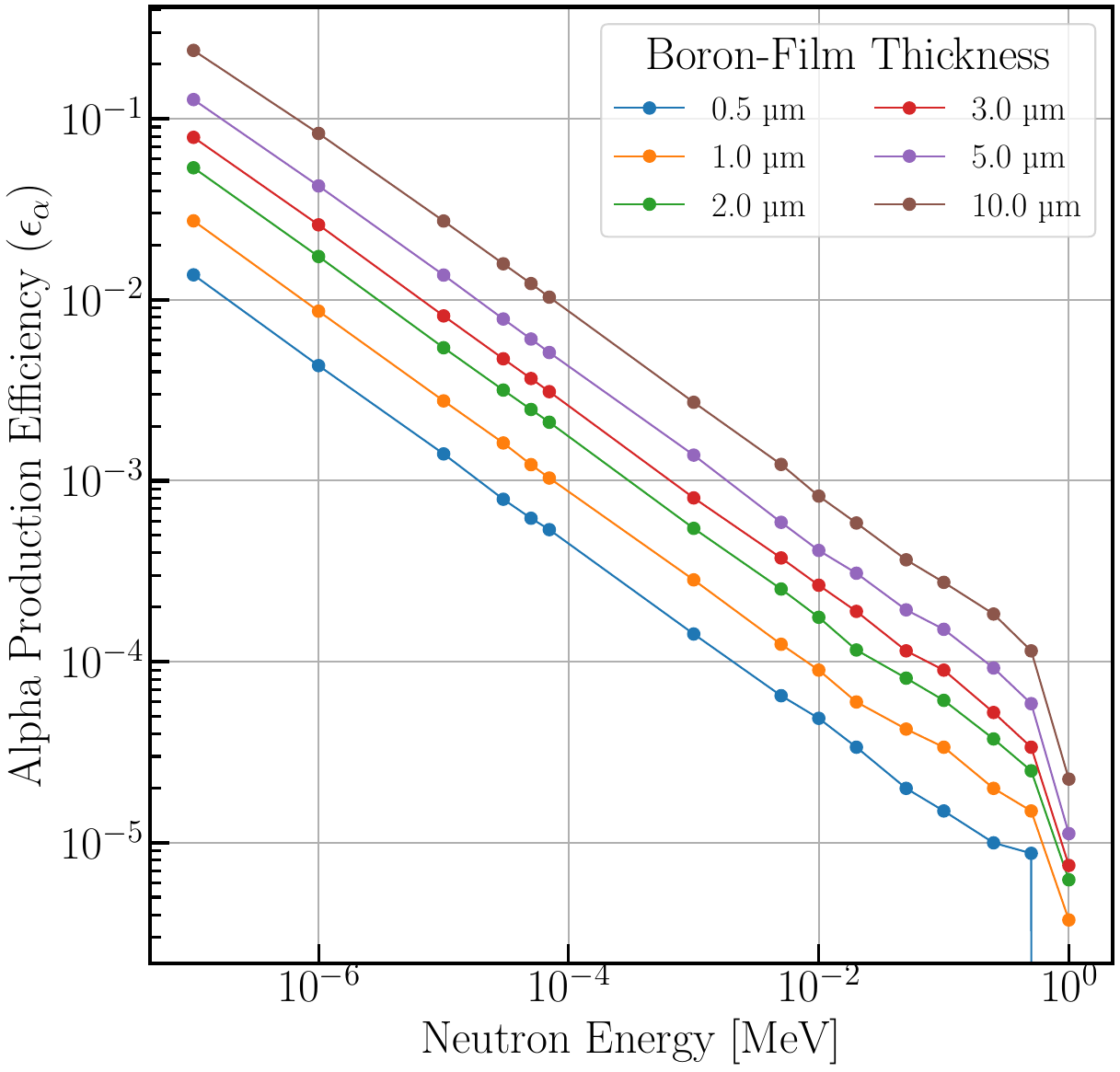}
    \caption{}
    \label{fig:captureEfficiency}
  \end{subfigure}
  \hfill
  % Second subfigure
  \begin{subfigure}[b]{0.475\linewidth}
    \centering
    \includegraphics[width=0.9\linewidth]{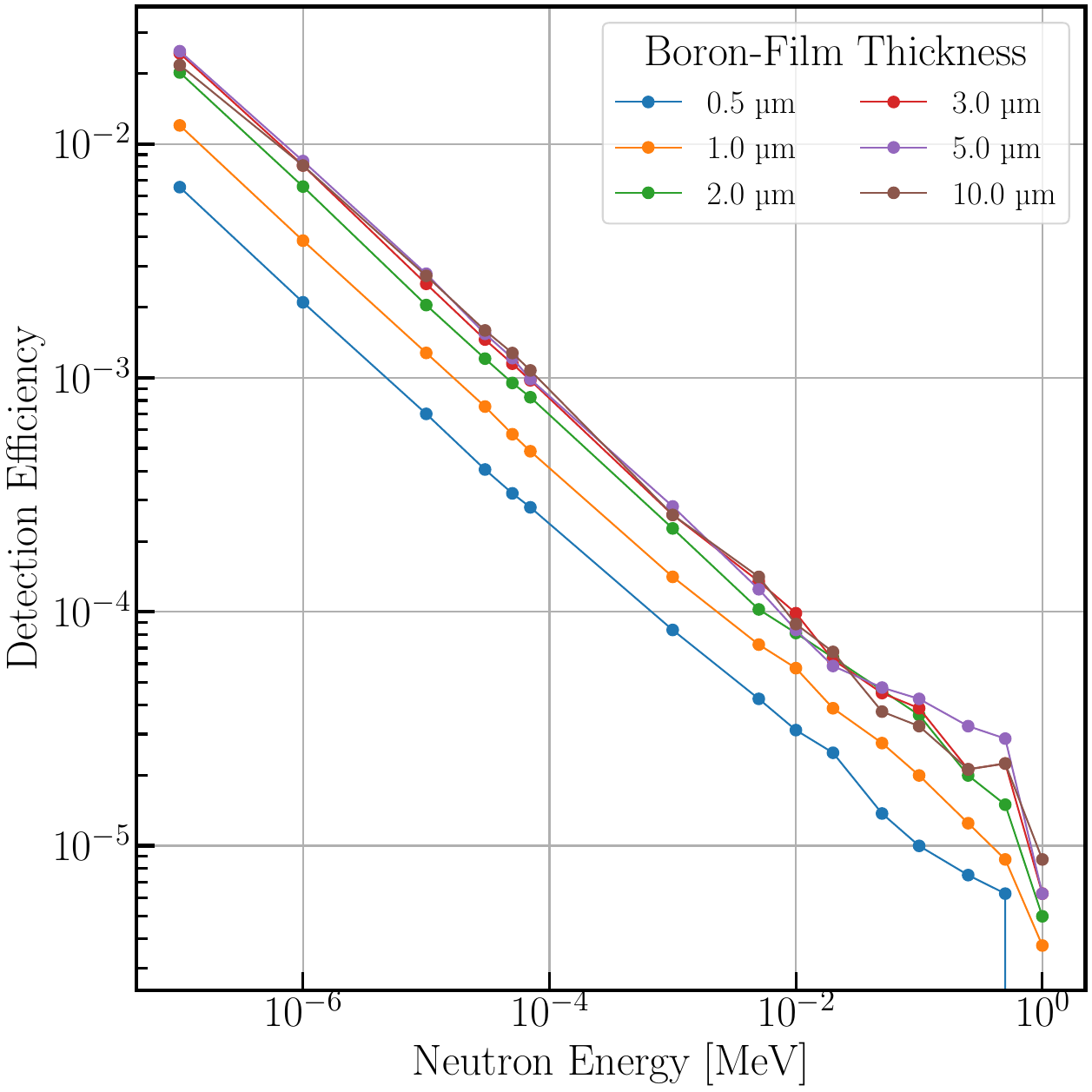}
    \caption{}
    \label{fig:detectionEfficiency}
  \end{subfigure}

   \caption{\textbf{a)} Efficiency of neutron capture (alpha production) in $^{10}$B films of different thickness as a function of neutron energy. \textbf{b)} The neutron detection efficiency of a silicon sensor with varying thicknesses of the \(^{10}\)B film. As discussed in the text, there isn't any notable increase in detection efficiency for films thicker than 3~$\mu$m, which is attributed to the limited range of the alpha particles. }
  \label{fig:sidebyside}
\end{figure}

To determine the alpha production in $^{10}$B and the corresponding alpha detection efficiency in silicon, we produced $10^9$ isotropic and monoenergetic neutrons 10\,cm from the geometric center of a boron-coated silicon sensor at energies ranging from $10^{-7}$\,MeV to 1\,MeV, and varied the boron-film thickness from 0.5\,$\mu$m to 10\,$\mu$m. The boron film and silicon sensor (1\,cm $\times$ 1\,cm $\times$ 750\,$\mu$m) were isolated in void to mimic space operation. In Figure~\ref{fig:captureEfficiency}, we present the neutron capture efficiency in the boron film through alpha production ($\epsilon_\alpha = [N_\alpha/N_n]_B$)\footnote{In this study, $N_i$ corresponds to the population of particle species $i$ in the material of interest.}. As expected, there is a linear gain in neutron capture efficiency as the boron film thickness is increased. However, after capture, the resulting alpha particles (or \(^{7}\)Li nuclei) must reach the silicon sensor to produce ionization, which determines the overall detection efficiency for neutrons of varying energies. In Figure~\ref{fig:detectionEfficiency}, we find that the alpha detection efficiency increases up to a boron film thickness of $\sim$3\,$\mu$m. For film thicknesses $\gtrsim$3\,$\mu$m, there are diminishing returns in alpha detection efficiency, indicating that the particles produced through neutron-capture terminate in the boron film before reaching the silicon. Therefore, we conclude that a 3\,$\mu$m-thick boron conversion layer is optimal to maximize neutron detection in a silicon sensor.

\subsubsection{Polyethylene Thickness Selection}

\begin{figure}[htb!]
  \centering
  % First subfigure
  \begin{subfigure}[b]{0.49\linewidth}
    \centering
    \includegraphics[width=0.9\linewidth]{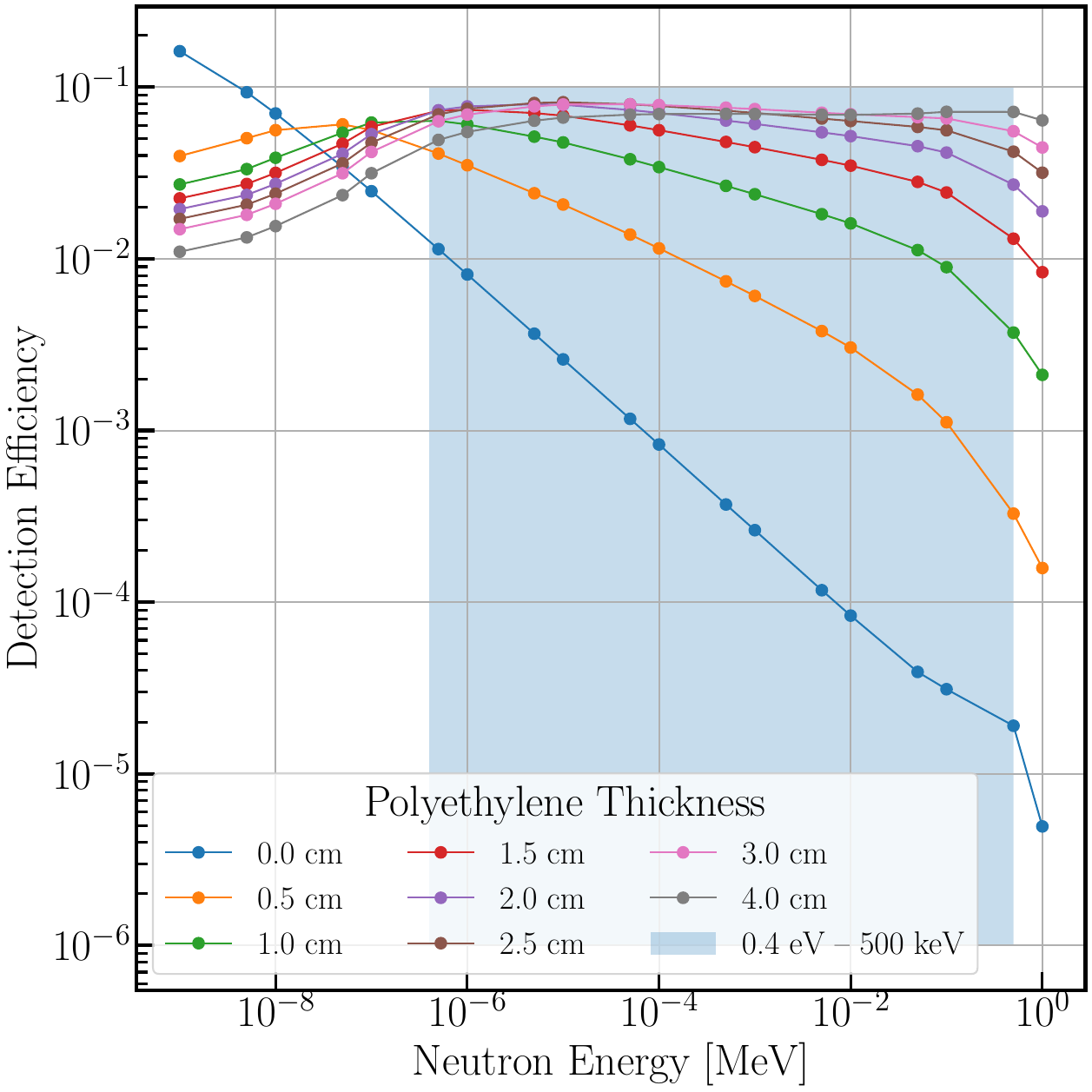}
    \caption{}
    \label{fig:detectionEfficiencyPoly}
  \end{subfigure}
  \hfill
  % Second subfigure
  \begin{subfigure}[b]{0.49\linewidth}
    \centering
    \includegraphics[width=0.9\linewidth]{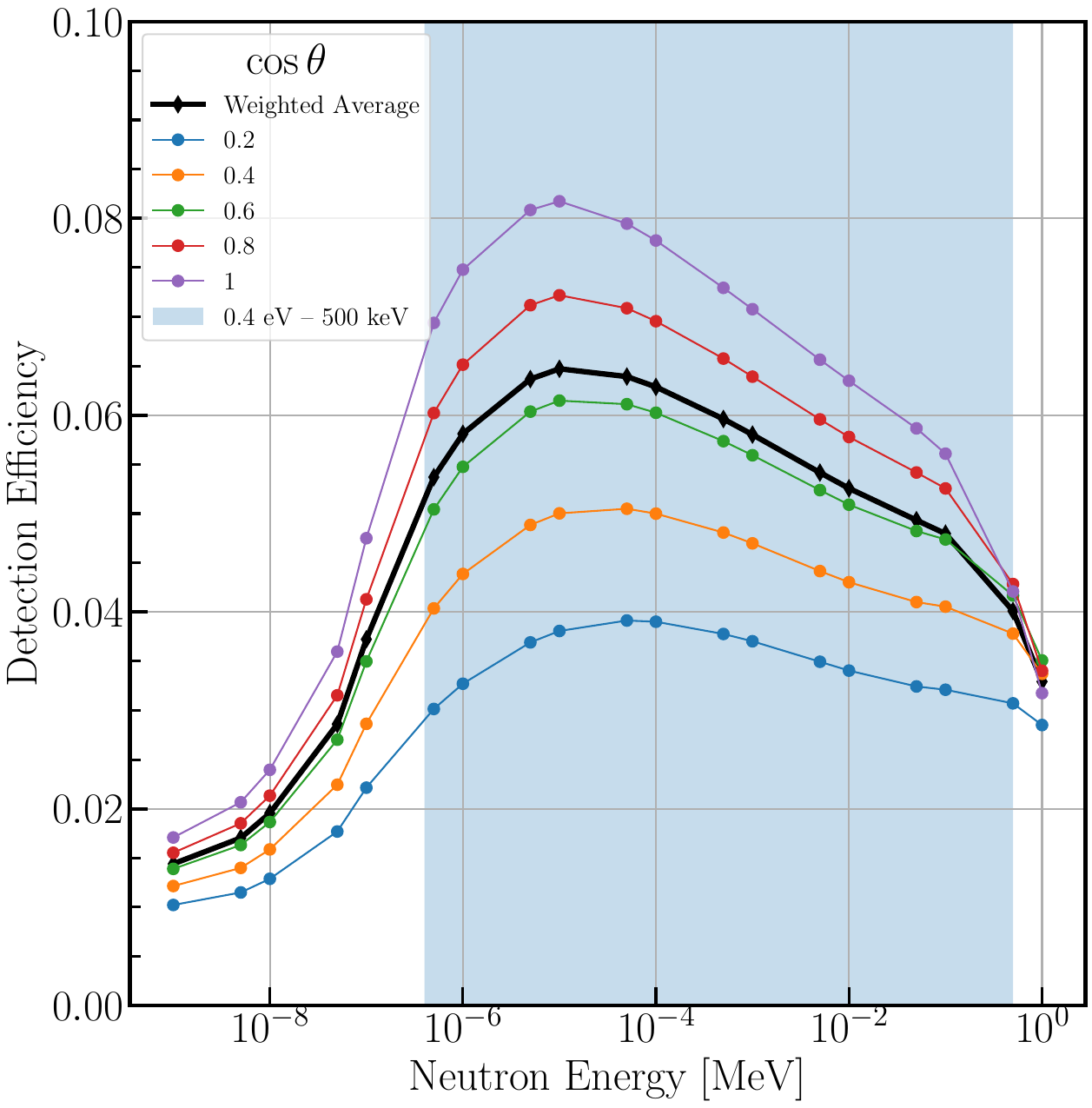}
    \caption{}
    \label{fig:detectionEfficiencyAngle}
  \end{subfigure}

  \caption{\textbf{a)} Neutron detection efficiency for a silicon sensor with a 3~$\mu$m \(^{10}\)B film inside a polyethylene moderator of varying thickness. Neutrons with energy below $\lesssim$1 eV are  preferentiably absorbed by the moderator, while neutrons with energy $\gtrsim$1 eV are slowed down in the moderator, increasing their detection efficiency. \textbf{b)} Detection efficiency as a function of incident angle $\cos{\theta}$. The final detector efficiency (black) is calculated as a weighted average of these curves, following the $\sqrt{\cos{\theta}}$  distribution in Figure \ref{fig:n_water_angles}.}
  \label{fig:sidebyside}
\end{figure}

To enhance the detection efficiency of epithermal neutrons in the boron-coated silicon detector, we introduce a polyethylene shield around the detector to moderate inbound neutrons. To select the optimal polyethylene thickness, we use a 750\,$\mu$m-thick detector with a 100\,cm $\times$ 100\,cm cross-sectional area, coated with a 3\,$\mu$m layer of $^{10}$B. The detector is encased in high-density polyethylene (HDPE), to moderate incoming neutrons. All materials are simulated in a void environment, as in Sec.~\ref{sec:filmthickness}. A series of simulations were conducted in MCNPX to study how HDPE thickness affects the detection of epithermal neutrons. In our simulations, we direct a pencil beam of 10$^8$ monoenergetic neutrons towards the geometric center of our detector and measure the detection efficiency of alpha particles in the silicon per source neutron, such that the detection efficiency of epithermal neutrons in our silicon detector is given by the relation $\epsilon_{det} = \left[ \frac{N_\alpha}{N_n}\right]_{Si}$ where $N_n = 10^8$. For our study, we vary the thickness of HDPE from [0.0, 4.0] cm, and the energy of the neutrons is varied between [10$^{-9}$, 1] MeV. In Figure~\ref{fig:detectionEfficiencyPoly} we note an improvement (up to 150\%) in detecting epithermal neutrons (represented in the shaded region) with 0.5\,cm of HDPE. At increasing thickness, the detection efficiency improves uniformly across all epithermal neutron energies, and we achieve a detection efficiency 
between 5-10\% of normally incident neutrons in our detector system\footnote{The detector system refers to the silicon detector coated with a 3\,$\mu$m boron film and encased in polyethylene. For simplicity, this system will be referred to as just `detector' hereafter.}.

The results presented in Figure~\ref{fig:detectionEfficiencyPoly} pertain to neutrons that are normally-incident on the detector surface. However, as illustrated in Figure~\ref{fig:n_water_angles}, we anticipate an angular distribution following $\sqrt{\cos{\theta}}$ for the neutrons striking the detector (with $\cos{\theta} = 1$ indicating a normal incidence). We simulated
the detector's energy-dependent neutron detection efficiency for a polyethylene thickness of 2.5\,cm for a range of incident angles, and the results are shown in Figure~\ref{fig:detectionEfficiencyAngle}. The overall detector efficiency is determined by weighing the angle-dependent results with the expected $\sqrt{\cos{\theta}}$ distribution of epithermal neutrons emerging from the lunar surface. This efficiency curve represents the final detector efficiency for this study, and is shown as the black curve in Figure~\ref{fig:detectionEfficiencyAngle}.

We note that our present simulation results only consider alphas in silicon; however, in the capture reaction in $^{10}$B, there is not a preferred direction of daughter emission, and $^7$Li ions are also produced. These ions are emitted in the opposite direction of an emitted alpha, and we can expect a corresponding increase in detection efficiency in the silicon from the $^7$Li ions since the ions have a similar range in boron as the alphas. This work provides a proof-of-concept that a boron-coated silicon detector encased in polyethylene can serve as an efficient epithermal neutron detector. Moreover, as discussed in Section~\ref{sec:MoonWaterNeutrons}, the addition of a thin Cd filter will further suppress neutrons with an energy $\lesssim$1 eV and enhance the detector's sensitivity to epithermal neutrons. In the next section, we will calculate the expected signal in a real detector at varying exposure times.

\subsection{Detector Sensitivity to Water}

The rate of epithermal neutrons detected by a boron-coated silicon sensor inside a polyethylene shield is
\begin{equation}
C_{\rm{epi}}(w=0)=\int_{0.4~eV}^{500 keV}\rm{d}E_n \epsilon(E_n)\times R(E_n)\times A_{det},
\end{equation}
where $A_{det}$  is the detector area, $\epsilon(E_n)$ is the detection efficiency in Figure~\ref{fig:detectionEfficiencyAngle}, and $R(E_n)$ is the expected neutron flux shown in Figure~\ref{fig:neutronfluxmoon}. For a 3\,$\mu m$ thick $^{10}$B film, 2.5\,cm thick moderator, and $A_{det}=10$\,cm~$^2$, we find $C_{\rm{epi}}(w=0)=2.7$\,cps.

\begin{figure}[ht!] 
    \centering
    \includegraphics[width=0.7\linewidth]{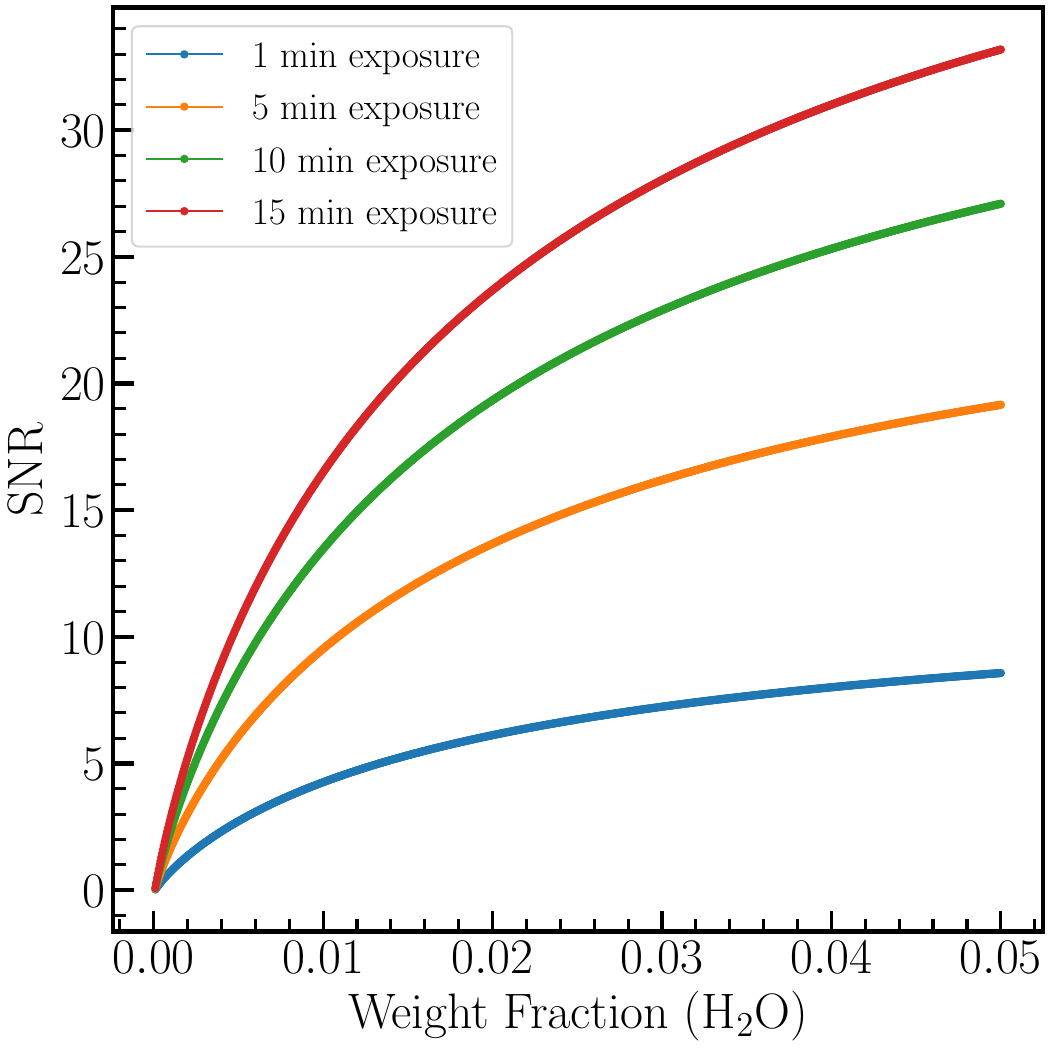}
    \caption{From Eq. \eqref{eqn:snr} we estimate the SNR associated with a 10\,cm$^2$ sensor encased in 2.5\,cm of HDPE at various exposure times. These results demonstrate the efficacy of our detector for detecting epithermal neutrons on the lunar surface.}
    \label{fig:sensitivity}
\end{figure}

The sensitivity for detection of H$_2$O under the lunar surface is calculated assuming the neutron flux dependence in Eq.~\eqref{eq:nfluxwwater}. The Signal-to-Noise-Ratio (SNR) is
\begin{equation} \label{eqn:snr}
    \mathrm{SNR}(w,T) = \frac{C_{\rm epi}(w)-C_{\rm epi}(w=0)}{ \sqrt{C_{\rm epi}(w=0)}} \sqrt{T},
\end{equation}
where $w$ is the H$_2$O weight fraction, and $T$ is the sample collection time (exposure time). $\mathrm{SNR}(w,T)$ is shown in Figure~\ref{fig:sensitivity}. The results indicate that a 5 minute exposure will produce $\mathrm{SNR}>8$ for H$_2$O weight fractions of $\sim$0.01.

\section{Water Detection in Lunar Rovers}
\label{sec:rover}

Over the next decade, a diverse set of national and commercial missions will deploy rovers to the lunar surface, with a strong focus on the south polar regions. NASA’s Artemis program is driving the development of the Lunar Terrain Vehicle (LTV) and the Endurance rover, designed to support both crewed operations and long-range autonomous exploration \cite{NASA_Artemis,NASA_LTV}. Commercial partners such as Lunar Outpost, Astrolab, and Intuitive Machines are delivering smaller robotic rovers under the Commercial Lunar Payload Services (CLPS) initiative, some of which are already operational. In parallel, China’s Chang’e~8, India–Japan’s LUPEX mission, and the International Lunar Research Station (ILRS) framework will contribute their own rovers, with timelines spanning from 2025 to the early 2030s~\cite{Change8_CNSA,LUPEX_JAXA,ILRS_CNSA}.
The primary scientific and strategic motivation behind many of these rover missions is the search for and utilization of lunar water ice, particularly in PSRs.

Equipping these upcoming rovers with a compact, high-sensitivity water detector—for example, an epithermal neutron sensor based on boron-coated silicon detectors—would directly address this knowledge gap. This detector technology provides a low-power alternative to traditional helium‑3 or scintillator-based instruments that is well-suited for small commercial and international rovers. The mobility of Artemis rovers further amplifies the value of in-situ water detection. The LTV is expected to travel at up to 15~km/h, meaning that in just 15 minutes it could traverse nearly 4~km of terrain~\cite{NASA_Teleops2024}. A sensor capable of detecting hydrogen at the 0.01~wt\% H$_2$O level would allow the rover to rapidly map water distribution over extended areas during short traverses.  This capability would be transformative for prospecting accessible ice and providing ground truth to orbital measurements, enabling informed site selection for resource extraction and long-term human presence.

Through targeted government–industry initiatives such as NASA’s Tipping Point program, rover design concepts have matured to the point where they can serve as practical baselines for payload sizing and integration. While the growing diversity of platforms emerging from these efforts offer many potential hosts for compact, low-mass payloads, the following section presents a representative baseline integration case.

\section{Instrument Packaging Concept}
\label{sec:lowpower}

The silicon-based detector concept described above, can be implemented in several ways, such as using an array of photodiodes~\cite{hamamatsuDiodes}, a CMOS or CCD imager, or a custom-designed Monolithic Active Pixel Sensor (MAPS)~\cite{MAPSHEP}. A key principle of this design is that when alpha particles (resulting from neutron capture) pass through a thin depleted silicon volume, they will be completely stopped via ionization losses after traveling just a few microns. In contrast, other particles, like protons, will pass through without fully stopping, allowing for effective discrimination between particle types in a space environment. Additionally, silicon sensors offer the advantages of low power consumption and lightweight construction, making them suitable to meet the requirements for deploying in a lunar rover.

While Artemis-class LTV rovers represent the upper bound of mobility and payload capacity for lunar surface exploration, we present here a detector benchmarked against a compact, commercially-developed rover class that can be delivered as a secondary payload on CLPS-type rovers. A representative example is the Astrobotic CubeRover family, which uses a standardized platform architecture to accommodate payloads in the 2-6\,U form factor (approximately 1\,kg per payload volume unit). CubeRovers achieve nominal traverse speeds of near 10\,cm\,s$^{-1}$ with kilometer-scale range over the mission life and are configured for short-cadence lunar-day operations~\cite{astrobotic_cuberover}. These platforms are intended to carry self-contained payloads with minimal interface requirements, providing a representative baseline for sizing and packaging a low-mass, low-power neutron detector.

To provide a specific solution that meets the requirements for a rover payload, we propose using a scientific CMOS sensor. As a benchmark, we consider a Teledyne CIS-120 back-illuminated CMOS sensor, which is commercially available off-the-shelf and manufactured using a space-qualified process. This sensor would feature a polyethylene moderator integrated as a structural casing with a thickness of 2.5\,cm, as found in Section~\ref{sec:sims}. Additionally, a 1\,mm thick cadmium filter will be included to attenuate low-energy neutrons.

The generalized system schematic is illustrated in Figure~\ref{fig:neutroncapture}. This 2048 × 4096 pixel sensor includes an on-chip timing sequencer and column-parallel ADCs to digitize the readout via LVDS outputs, with configuration accomplished through an SPI interface. A compact image processing and power conditioning board will support the operation of the sensor, enabling on-board data reduction to extract metadata (neutron counts) while discarding raw frames.

An initial packaging concept for the boron-coated silicon neutron detector was sized for a CubeRover-class platform to assess the package sizing within representative platform constraints. As outlined in Table~\ref{tab:baseline}, the assembly mass breakdown of the key detector components occupies a small fraction of the allowable mass capacity for the 4U CubeRover variant. Continuous power consumption is estimated at $\sim$1.64\,W, within the 2\,W allocation implied by the 0.5\,W\,kg$^{-1}$ guideline~\cite{astrobotic_cuberover}. The estimated science data rate for a 10-minute integration cadence remains well below CubeRover-class payload uplink limits. At a traverse speed of 10\,cm\,s$^{-1}$, a 10-minute integration bin corresponds to approximately 60\,m of ground track, providing opportunities for hydrogen mapping in chained cadences.

For continuous operation intervals spanning a full lunar day (approximately 14 Earth days), the data budget for a 10-minute cadence is given in Table~\ref{tab:baseline}. Each record assumes 16-bit resolution and includes a 30\% overhead for timestamps and packet encapsulation, yielding approximately 21.5\,MB over a Lunar day. The data generation is sized for the CubeRover’s 4\,GB onboard payload storage and 40\,kbps data throughput rate, with substantial margin for system development contingencies and scaling opportunities. The spatial distribution of epithermal neutron counts is intended to complement co-payload or independent lunar surface context data, such as topography or compositional maps, for cross-correlation analyses.

Scalable elements such as the sensor, moderator, and filter support are compatible with other small-rover payloads, and can be integrated without altering the basic detection elements. The present configuration fits within a standard 1U envelope, leaving margin for integration hardware. Larger rovers in the CubeRover family (such as the 6U variant) can carry multiple modules that increase the active detection area. Within these bounds, the packaging concept remains adaptable to resource-constrained lunar-day missions while maintaining compatibility with the system-level interfaces of CubeRover-class hosts.

\begin{table}[htbp]
\centering
\caption{Detector package parameters compared to CubeRover\,4U baseline constraints}
\label{tab:baseline}
\renewcommand{\arraystretch}{1.15}
\setlength{\tabcolsep}{8pt}
\footnotesize
\begin{tabularx}{\textwidth}{
  >{\raggedright\arraybackslash}p{0.18\textwidth}  % Section header
  >{\raggedright\arraybackslash}X                  % Parameter
  >{\centering\arraybackslash}p{0.18\textwidth}    % Value (centered)
}
\toprule
\textbf{Section} & \textbf{Parameter} & \textbf{Value} \\
\midrule

\multirow[t]{6}{*}{\textit{Mass [g]}}%
  & Silicon Sensor & 20 \\
  & Polyethylene moderator & 49 \\
  & Cadmium filter & 15 \\
  & Image processing electronics & 70 \\
  & \textit{Total mass with 50\% margin} & 231 \\
  & \textbf{Baseline mass$^{\dagger}$} & \textbf{4000} \\
\hline
\addlinespace
\multirow[t]{4}{*}{\textit{Continuous Power [W]}}%
  & Silicon Sensor & < 0.5 \\
  & Readout Electronics & < 1.0 \\
  & \textit{Total detector power} & $\sim1.5$ \\
  & \textbf{Baseline power$^{\dagger}$} & \textbf{2.0} \\
\hline
\addlinespace
\multirow[t]{8}{*}{\textit{Data [MB]}}%
  & Traverse speed & 10\,cm\,s$^{-1}$ \\
  & Integration cadence & 10\,min exposures, on-board binning \\
  & Exposures per 24-hour day & 144 \\
  & Metadata per exposure (16-bit resolution; 30\% overhead) & 0.01 \\
  & Data per 24-hour day & 1.50 \\
  & \textit{Data over Lunar day} & 21.5 \\
  & \textbf{Baseline storage$^{\dagger}$} & \textbf{4000} \\
\bottomrule
\end{tabularx}
\par\smallskip
{\footnotesize $^{\dagger}$\,Baseline constraints from CubeRover 4U payload accommodation (Astrobotic)\cite{astrobotic_cuberover}.}
\end{table}

\section{Conclusion}
\label{sec:conclusion}

The detection of water is essential for both current and future lunar exploration plans. Several past missions have conducted surveys to locate water on the moon's surface using infrared spectroscopy~\cite{IRLUNARMAP}. Additionally, previous lunar orbiters have investigated the presence of underground water by measuring the rate of epithermal neutrons with helium-3 detectors \cite{Mitrofanov2010_LEND}. Recent developments in solid-state neutron imaging technology, based on neutron capture in $^{10}$B~\cite{Blostein2014,Kuk_2021}, offer an alternative method for monitoring the rate of epithermal neutrons in the lunar environment.

In this work, we described a conceptual design for an epithermal neutron counter based on a silicon imager with a thin conversion layer of $^{10}$B and a polyethylene neutron moderator. This detector is designed to monitor epithermal neutrons on the lunar surface. Our results demonstrate that we can achieve an efficiency of 7\% for 0.4\,eV-500\,keV neutrons using this technology. The simulations indicate that the rate of epithermal neutrons detected with a 10\,cm$^2$ detector is $\sim$3 counts per second. Given that the flux of epithermal neutrons is dependent on the underground water content (see Eq.~\eqref{eq:nfluxwwater}), this low-mass and low-power detector could provide a high significance ($\mathrm{SNR}> 5$) signal for a  H$_2$O weight ratio of 0.01\% with just a 5-minute exposure time. The low-mass and low-power characteristics of this detector make it suitable for deployment on future lunar rovers, enabling searches for underground water without the need for excavation.

\section{Acknowledgement}
\label{sec:acknowledgement}

This work was done using the resources of the Fermi National Accelerator Laboratory (Fermilab), a U.S. Department of Energy, Office of Science, Office of High Energy Physics HEP User Facility. Fermilab is managed by FermiForward Discovery Group, LLC, acting under Contract No. 89243024CSC000002. This work was supported in part by the Department of Astronomy and Astrophysics at the University of Chicago.

\clearpage
\newpage
\addtocontents{toc}{\protect\setcounter{tocdepth}{1}}

\section*{}\label{sec:references}\addcontentsline{toc}{section}{References}
\bibliographystyle{unsrt}
\bibliography{main,main-luna}
\clearpage

\end{document}